%% file: main.tex
\documentclass[sigconf,nonacm,screen]{acmart}
\renewcommand\footnotetextcopyrightpermission[1]{}
\AtBeginDocument{%
  }

\usepackage{import}
\usepackage{xcolor}
\usepackage{array}

\begin{document}

\title{VizCopilot: Fostering Appropriate Reliance on Enterprise AI Chatbots with Context Visualization}


\author{Sam Yu-Te lee}
\affiliation{%
  \institution{University of California, Davis}
  \city{Davis}
  \country{United States}}
\email{ytlee@ucdavis.edu}

\author{Jingya Chen}
\affiliation{%
  \institution{Microsoft}
  \city{Redmond}
  \country{United States}}
\email{jingyachen@microsoft.com}

\author{Albert Calzaretto}
\affiliation{%
  \institution{Microsoft}
  \city{Redmond}
  \country{United States}}
\email{acalzaretto@microsoft.com}

\author{Richard Lee}
\affiliation{%
  \institution{Microsoft}
  \city{Redmond}
  \country{United States}}
\email{richalee@microsoft.com}

\author{Samir Passi}
\affiliation{%
  \institution{Microsoft}
  \city{Redmond}
  \country{United States}}
\email{sp966@cornell.edu}

\author{Alice Ferng}
\affiliation{%
  \institution{Microsoft}
  \city{Redmond}
  \country{United States}}
\email{Alice.Ferng@microsoft.com}

\author{Mihaela Vorvoreanu}
\affiliation{%
  \institution{Microsoft}
  \city{Redmond}
  \country{United States}}
\email{Mihaela.Vorvoreanu@microsoft.com}


\begin{abstract}
Enterprise AI chatbots show promise in supporting knowledge workers in information synthesis tasks by retrieving context from large, heterogeneous databases before generating answers. However, when the retrieved context misaligns with user intentions, these chatbots often produce ``irrelevantly right'' responses that raise the risk of overreliance. The current AI interaction paradigm asks users to verify chatbot outputs, without providing sufficient transparency and control over context. We propose and explore the value of a different approach that gives users control over context engineering through data visualization. We conduct a Research-through-Design study using VIzCopilot, a research prototype that extends an AI chatbot with corpus visualization techniques. We evaluate the potential of this approach through a qualitative user study that compares VizCopilot with a chat-only AI interface. Our findings show that visualization helps users detect and correct misaligned context, while also enhancing user agency and system transparency. This work demonstrates the viability of alternate interaction approaches to fostering appropriate reliance on AI and surfaces design opportunities for enterprise AI chatbots. At the same time, the study reveals limitations regarding verification support and trust in AI summaries. Based on these findings, we discuss how visualization can foster appropriate reliance, the associated design trade-offs, and our vision towards transparent and sustainable human-AI interaction.

\end{abstract}




\makeatletter
\def\input@path{{sections/}}
\graphicspath{{figs/}{figures/}{images/}{./}} 
\makeatother

\definecolor{oiSkyBlue}{RGB}{86,180,233}
\definecolor{oiBluishGreen}{RGB}{0,158,115}
\definecolor{oiOrange}{RGB}{230,159,0}

\newcommand{\theme}[2]{%
  \paragraph{\textbf{Theme #1: #2.}}%
}
\maketitle

\input{01_introduction.tex}
\input{02_related_works}
\input{03_design_analysis}
\input{04_methodology}
\input{05_user_study}

\input{06_results}

\input{07_discussion}

\section{Generative AI Usage}
The authors used generative AI tools solely for proofreading purposes, such as correcting grammatical errors or typos. No generative AI was used to generate technical content, analyses, results, or conclusions in this paper.

\bibliographystyle{ACM-Reference-Format}
\bibliography{references}


\end{document}

%% file: sections/01_introduction.tex

\section{Introduction}
Enterprise AI chatbots such as Microsoft 365 (M365) Copilot~\cite{Microsoft_M365_Copilot_2023} can retrieve data from large, heterogeneous enterprise databases. These systems promise to support users in information synthesis by condensing immense enterprise data into digestible passages~\cite{lewis2020rag}.
However, user studies on such chatbots suggest that users often receive responses that are ``irrelevantly right'': plausible answers that fail to address the user’s actual intent~\cite{microsoft2025copilot_moat, subramonyam2024gulf_of_envisioning}. Users accepting plausible but incorrect answers is a problem known as overreliance on AI~\cite{passi2022overreliance}. Overreliance poses a serious risk in the enterprise as it can silently misguide decision-making. It has garnered attention from both socio-technical researchers and public sectors~\cite{EU_AI_Act, DEC_AI_Literacy}.

For example, the EU AI Act~\cite{EU_AI_Act} mandates that high-risk AI systems be designed with effective human oversight, such that users can understand system outputs, detect erroneous or misleading recommendations, and intervene or override decisions when necessary.
Socio-technical researchers have begun to investigate strategies that mitigate overreliance~\cite{buccinca2021cognitiveforcing, drosos2025itmakesthinkprovocations, passi2024appropriate, Kim2024uncertainty_expression} and proposed guidelines~\cite{amershi2019guidelineHAI, Microsoft_2025_OverrelianceAI} for UX designs.
Together, these efforts reflect the urgency of fostering appropriate reliance on AI.

In enterprise AI chatbots, the issue of producing ``irrelevantly right'' outputs stems from the data retrieval process, which is increasingly referred to as context engineering: ``in every industrial-strength LLM application, context engineering is the delicate art and science of filling the context window with just the right information for the next step.''~\cite{Karpathy2023Context}. 
Research in context engineering suggests that autonomous retrieval approaches are susceptible to \textbf{context misalignment}, which typically manifests in two key issues. First, the retrieved context may be irrelevant to the user’s prompt. Current retrieval methods are searching algorithms based on semantic vectors, keywords, and metadata such as timestamps or file types~\cite{mei2025surveycontextengineering}. These algorithms might struggle to adapt to the messy formats, deprecated files, incorrect metadata, and out-of-context terminologies often seen in enterprise databases~\cite{mukherjee2004enterprise}. 
Second, the chatbot may synthesize the retrieved context inappropriately due to a lack of background knowledge, incorrect assumptions, misinterpretations, or limited reasoning ability. 
In light of these issues, we seek to explore whether giving users more control over context engineering can be a viable strategy for producing AI outputs that are better aligned with user goals, and for fostering appropriate reliance on AI.
As such, we are interested not only in alerting users of potentially incorrect responses, but also in effective interaction mechanisms that allow users to modify context and steer responses during regular interaction~\cite{He2026AIchart_search}. 

A user interaction mechanism that would enable context engineering poses design challenges as it would require users to examine the retrieved context carefully and thoroughly, which is a significant cognitive load.
To address this challenge, we took inspiration from corpus visualization research that supports large-scale information sensemaking and synthesis with interactive and progressive visualizations, and employed similar techniques to minimize cognitive load~\cite{liu2019textvis}. 
Our design rationale is that enabling users to visually search and navigate contextual data can (1) surface misalignment between retrieval and user intent, and (2) help users develop a sufficient understanding of the context to identify and correct problematic AI synthesis, thus fostering more appropriate reliance on AI.

Based on this rationale, we developed VizCopilot, a system that extends a chat-based enterprise application such as M365 Copilot with a treemap-based context visualization, as a design probe. We explored the value of giving users control of context engineering by conducting a Research-through-Design (RtD) study with 14 participants in a within-subject setting. Participants compared VizCopilot with a simplified pure-text AI chatbot in performing information synthesis tasks on a synthetic dataset.
Results show that our approach helps users correct misaligned context and adapt prompting strategies for better context retrieval and enhances transparency and confidence, despite limitations regarding verification support and trust in AI-generated summaries.
Based on these insights, we discuss the potential for visualization to help foster appropriate reliance through increased transparency and user control, design trade-offs regarding usability, as well as a vision to facilitate a sustainable human-AI collaboration paradigm. 

\noindent This work makes the following contributions:
\begin{itemize}
    \item We explore a context engineering approach that supports human oversight and intervention through data visualization, and evaluate its potential for fostering appropriate reliance.    
    \item We report findings from the user study, which shows the viability of incorporating visualization for context alignment, as well as the improved user confidence and sense of control. We also report limitations and future design directions for visualization-enhanced chatbots. 
\end{itemize}

%% file: sections/02_related_works.tex
\section{Related Work}
\subsection{Context Engineering for LLM Chatbots}
Context engineering~\cite{mei2025surveycontextengineering} is an emerging area concerned with supplying large language models (LLMs) with precise contextual information. Its foundation lies in retrieval-augmented generation (RAG)~\cite{lewis2020rag}, which was introduced to enhance factual accuracy in knowledge-intensive tasks. RAG and subsequent advances in context engineering typically rely on dense retrieval methods~\cite{karpukhin2020dpr} to identify relevant information units (e.g., documented facts) from external databases and incorporate them into the model's conversation history. This mechanism enables LLM-based chatbots to access up-to-date knowledge without retraining and further contributes to transparency, as the retrieved passages can function as verifiable evidence~\cite{Yu_2025ragsurvey}. 

While promising, context engineering is subject to several limitations that have been well documented in the literature. First, the retrieved context may be irrelevant or only partially relevant to the user’s query~\cite{karpukhin2020dpr}, i.e., the retrieved materials omit critical information necessary for producing a correct response~\cite{salemi2024evaluatingrag}. Second, retrieved documents may present inconsistent or even contradictory content~\cite{Yu_2025ragsurvey}, which can induce model biases or lead to inconsistent outputs. Third, models may err in synthesis even when the appropriate context is provided. For example, the model may hallucinate, misinterpret the retrieved evidence, or reason from inappropriate assumptions~\cite{huang2025LLMhallucinationsurvey}.
Motivated by these challenges, we examine the value of giving users direct control over context engineering through data visualization.

\subsection{Design Studies for Fostering Appropriate Reliance on LLM Chatbots}
Despite their widespread integration across application layers, LLM-based chatbots introduce new interaction design challenges, most notably overreliance, in which users accept incorrect or misleading AI outputs~\cite{passi2025addressing}.
Previous work on interface design has sought to mitigate this issue through various strategies. Some studies focus on providing explanations that assist in verification and conveying uncertainty through highlights or linguistic expressions~\cite{passi2024appropriate, Kim2024uncertainty_expression}. Others take a more radical approach, introducing cognitive forcing functions~\cite{buccinca2021cognitiveforcing} that deliberately interrupt routine workflows. These functions use session timeouts or short textual alerts alongside AI responses to highlight risks, limitations, and alternatives, thereby provoking critical reflection~\cite{drosos2025itmakesthinkprovocations}.

All these interventions aim to raise awareness of chatbot failures. In practice, users may lack the experience or cognitive capacity to manually evaluate lengthy responses, cross-check supporting evidence, and translate such evaluations into actionable prompt refinements~\cite{zamfirescu2023whyjohnny, jiang2022promptmaker, kim2023evallm}.
This gap reflects a broader design challenge: many chatbot interfaces implicitly demand sustained metacognitive effort, such as monitoring context, assessing reliability, and planning corrective actions, that exceeds what users can reasonably maintain during everyday work~\cite{tankelevitch2024metacognition}.

In this work, we take an approach different than raising awareness of chatbot failures. Our goal is to explore context engineering as a strategy that enables users to effectively correct errors and steer chatbot behavior. We design visualizations that afford offloading key cognitive and metacognitive activities, including tracking context, inspecting supporting evidence, and modifying retrieved context, onto visual representations. By shifting these demands from users' internal reasoning to shared visual structures, our approach aims to augment, rather than replace, human judgment in human–AI interaction~\cite{heer2019sharedrepresentation, Ding2025diagram_guardrail}. 
By supporting users to engage in context engineering, this approach aims to foster appropriate reliance in two ways: 1) By actively reviewing and selecting context, users can develop better awareness, making it easier to catch and correct mistakes or `irrelevantly right'' chatbot responses; 2) With better context engineering, the accuracy of chatbot responses may improve.

%
%
\subsection{Information Sensemaking and Synthesis with Corpus Visualization}
Visualization has played a central role in supporting information sensemaking and synthesis by offloading cognitive effort onto persistent visual representations and by surfacing patterns that are difficult to detect through text alone~\cite{li2025role_vis_kg}.
Prior work has shown that effective visual sensemaking tools must balance exploration with interpretability, while supporting users’ ability to exercise judgment and assess the reliability of information~\cite{liu2019textvis}.

Early visualization systems for large corpora relied on topic modeling or keyword extraction techniques~\cite{alexander2014serendip, dou2013hierarchicaltopics}
However, constrained by the ``bag-of-words'' representations of most topic models, these approaches were criticized for having low interpretability~\cite{lee2017human_touch,chuang2012interpretation_and_trust}.
More recent work has shifted toward embedding-based topic modeling, typically using dimensionality reduction (DR) to project documents or keywords onto scatterplots~\cite{park2018conceptvector, narechania2022vitality, choo2013utopian}. 
By placing semantically similar items closer together, these systems reveal clusters that can suggest latent topical structure. However, such visualizations often introduce substantial complexity: their layouts can be sensitive to modeling and parameter choices, their axes lack semantic grounding, and dense projections can lead to visual clutter~\cite{jeon2025survey_DR}. As a result, effectively interpreting these representations can require considerable visualization literacy, limiting their suitability for efficient information seeking and validation by many knowledge workers~\cite{burns2023visualization_novice}.

In response, VizCopilot builds upon a treemap design familiar to most knowledge workers and extends it with decluttered DR scatterplots and progressive disclosure to better align with users’ needs. 
In addition, VizCopilot positions the chatbot as the central component, with visualization serving as a complementary means of offloading cognitive demands, catering to knowledge workers' existing workflows~\cite{bano2025qualitativestudyuserperception}.


%% file: sections/03_design_analysis.tex
\section{Design Analysis}
Drawing inspiration from the literature on corpus visualization~\cite{liu2019textvis} and context engineering~\cite{mei2025surveycontextengineering}, we analyze the design challenges that arise from the large-scale and complex nature of enterprise data. In the following section, we summarize these challenges and present a set of design requirements for our research prototype.

\paragraph{Challenges}
\textbf{First}, the retrieved context is typically too large to be meaningfully displayed. Conventional UI components such as paginated list views, provide limited support for sensemaking of such large-scale context. Expecting users to skim through multiple pages and synthesize the information is impractical~\cite{liu2019textvis}.
\textbf{Second}, users need to verify the context to decide how to steer it. However, users are typically either unaware of its necessity or reluctant to invest the required effort~\cite{vorvoreanu2025fostering}.
Even when users recognize the need, comprehensive verification still requires them to closely read the retrieved data items and cross-reference them with the response. This process is not only cognitively demanding but also undermines the primary benefit of enterprise chatbots, i.e., automating information synthesis~\cite{yun2025genAIknowledgework}. 
\textbf{Third}, modifying context file by file, as with standard file selection and upload controls, is both inefficient and largely ineffective. Small adjustments have minimal impact because the overall semantics of the retrieved context remain unchanged~\cite{Lee2025awesum}. Chatbots are unlikely to be sensitive to such fine-grained modifications within a large-scale context blob.

\paragraph{Design Requirements}
Considering these challenges, we summarize three design requirements (\textbf{DRs}):
\begin{itemize}
    \item \textbf{DR1: Sensemaking support.} Users need to develop enough knowledge of context to be able to steer the chatbot.
    To reduce cognitive load, corpus visualizations can scaffold this process by organizing the data with topics, keywords, or other metadata. When combined with progressive disclosure, such scaffolding allows users to navigate information incrementally.
    \item \textbf{DR2: Verification support.} Beyond sensemaking, the system should support users in verification by both raising awareness of the need to verify responses and by reducing the cognitive load of verifying LLM outputs against context data. Corpus visualizations can highlight areas of context that require verification and leverage progressive disclosure to minimize the amount of information users must process during verification.
    \item \textbf{DR3: Control at group-level.} Users need the ability to examine and modify context data at the group level rather than file-by-file, to ensure that changes are substantial enough to steer chatbots effectively. This capability can be supported through direct manipulation of visual elements in the corpus visualization. Progressive levels of information grouping should offer increasingly fine-grained yet semantically meaningful group-level selections, enabling context modification with efficiency and precision.

\end{itemize}


%% file: sections/04_methodology.tex
\begin{figure*}[t]
    \centering
    \includegraphics[width=\textwidth]{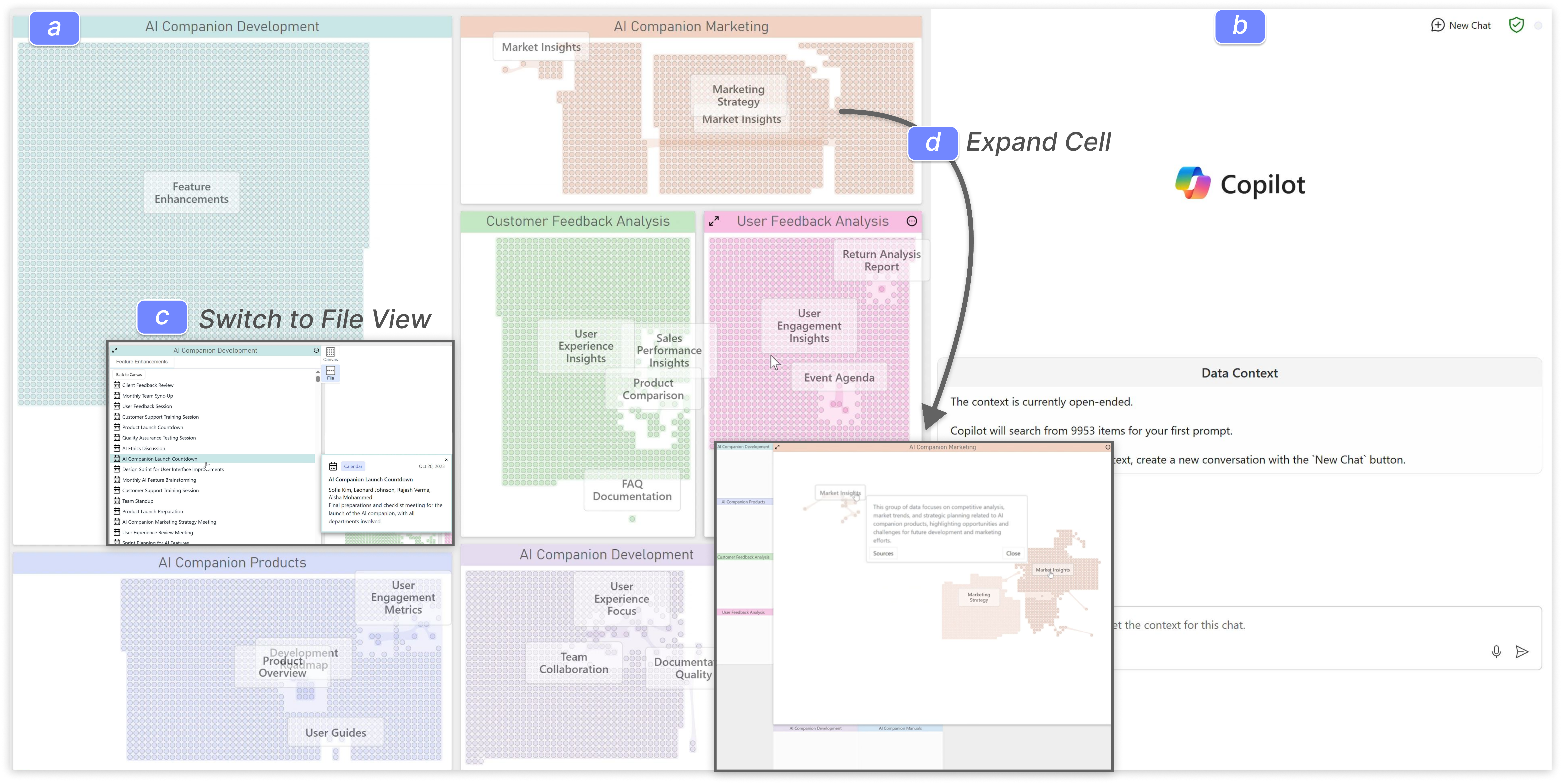}
    \caption{Overview of VizCopilot before entering a prompt. (a) The visualization panel shows topic structures of the context in a treemap-based visualization. (b) The Copilot chat panel allows for typical conversational interactions with an extension of the data context panel. (c) Each cell can be switched between the canvas view (default) and the file view, which allows for direct inspection of file content. (d) Each cell in the treemap can be expanded to allocate more space for visual clarity.}
    \Description{}
    \label{fig: overview}
\end{figure*}
\section{VizCopilot: Prototype Design}
Based on these design requirements, we developed our research prototype: VizCopilot, an interactive prototype that extends the typical chat interface with a context visualization panel. VizCopilot is not intended as a mature system; rather, it serves as a design probe that operationalizes the design requirements to examine the viability of giving users control over context engineering through data visualization and to surface opportunities for future design. In this section, we present the interface and data pipeline design and explain how they address the design requirements.
\subsection{Interface Design}
The interface of VizCopilot is divided into two main components (\autoref{fig: overview}): a visualization panel and a chat panel that is similar to a regular chatbot. 

\subsubsection{Extended Treemap Visualization}\label{sec: treemap}
The context data available to the chatbot is preprocessed using topic modeling techniques and presented through an extended treemap visualization. Each treemap cell represents a topic, with its area proportional to the number of data items it contains. Within each cell, additional subtopics are extracted and labeled to support sensemaking (\textit{DR1}). We opted for treemap because it is a common visualization used for databases that most enterprise chatbot users are familiar with. It also allows us to provide a topical structure personalized to the user’s database.  

Different from a conventional treemap, each data item within a cell is represented as a circle. 
The position of the circles is generated in two steps.
First, each circle is assigned an initial coordinate based on KernelPCA~\cite{scholkopf1997kernel} applied to the embedding of its textual content. Second, to reduce visual clutter, each treemap cell is partitioned into a grid according to its size and the number of items it contains. Data items are then assigned unique grid points to eliminate occlusion. The decluttering is ordered by subtopic to maintain clear boundaries.  

The extended treemap visualization essentially transforms the treemap, an aggregated visualization, into a unit visualization~\cite{park2018atom}. As discussed in previous research, unit visualizations are particularly well-suited for visualization novices, as the one-to-one mapping between data items and visual marks avoids any additional abstraction layers when interpreting the visualization. Moreover, this extension supports multiple useful purposes: it enables context to be highlighted in a way that supports volume estimation, (i.e., estimating how many items are highlighted in each cell), affording more intuitive selection interactions; it shows users not only what data are retrieved, but also what is not retrieved, which is critical information for context alignment; it maintains visual continuity in interactions, helping user to develop a ``data sense'' of the information available to them. 
To better reveal the topical structure, a cell can be expanded to fill the visualization panel while keeping surrounding cells interactable. Technical considerations regarding topic modeling and dimensionality reduction are detailed in~\autoref{sec: data_pipeline}.

\subsubsection{Coordinations}
\begin{figure*}[t]
    \centering
    \includegraphics[width=\textwidth]{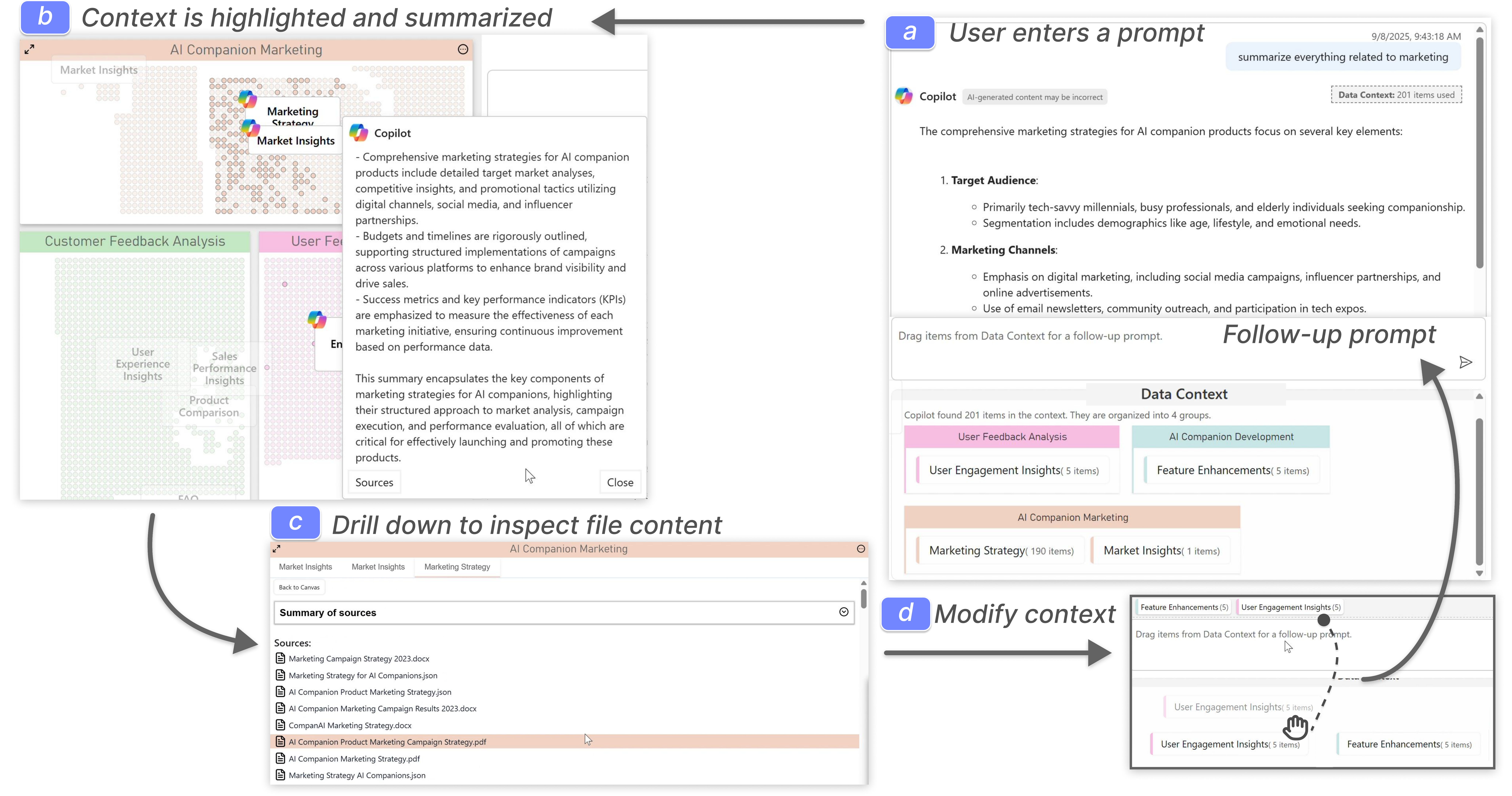}
    \caption{Interactions supported by VizCopilot. (a) Users can enter their prompt in the chat panel to initiate a conversation. (b) VizCopilot uses the prompt to retrieve context data, highlight it on the visualization, and automatically summarize it according to the subtopics. (c) Users can drill down to individual subtopics and inspect the file contents. (d) Users can use the drag-and-drop feature to modify context for follow-up prompts. }
    \Description{}
    \label{fig: interaction}
\end{figure*}

The visualization and chat panel are coordinated in several ways.
\paragraph{Highlighting retrieved context} 
The unit visualization design allows for highlighting individual items retrieved as context, as shown in~\autoref{fig: interaction}-b. The highlighting effect keeps the irrelevant items and subtopic boundaries visible to maintain visual continuity and support volume estimation (\textit{DR1}). 
Note that the expansion of cells and all relevant interactions are designed to be consistent when the highlight effect is active. 

\paragraph{Group-level Control}
The data context panel (\autoref{fig: overview}-d) presents thumbnails of subtopics with relevant data items. The panel uses common UI components and serves as a buffer zone for users who may feel intimidated by the unfamiliar designs in the visualization. To support this role, it is positioned adjacent to the chat panel and adopts a simple row-of-cards layout.
The data context panel supports two key interactions for group-level control (\textit{DR3}). First, clicking a subtopic thumbnail highlights the corresponding subtopic in the visualization, enabling users to quickly locate areas of interest without skimming the entire visualization. Second, users can drag and drop thumbnails into the chat panel, allowing them to specify a subset of context for the chatbot.

\paragraph{Progressive Disclosure}
The information in VizCopilot is organized across three levels. At the \textbf{highest} level, which has the broadest scope and lowest information density, the system provides a topical overview of the context and highlights retrieved context. This view allows users to quickly grasp the distribution of context across topics and detect potential context misalignments (\textit{DR2}). The \textbf{intermediate} level consists of AI summaries for individual subtopics (\autoref{fig: interaction}-b). These summaries provide greater detail on the relevant data items and explain their relevance to the prompt, reducing the cognitive effort (\textit{DR1}). At the most \textbf{granular} level, the file view allows users to select and inspect individual files, enabling close reading and in-depth sensemaking on the raw data.

\subsection{Data Pipeline}\label{sec: data_pipeline}
At preprocessing stage, the context dataset is processed with topic modeling and dimensionality reduction techniques to create topical scaffoldings.
 At runtime, the system employs a retrieval-augmented-generation (RAG) architecture extended with subtopic summaries to respond to user. 
During development and user study, VizCopilot uses a synthetic dataset that contains corporate data of a fictitious AI companion company. We present the algorithmic choices from a technical perspective and explain how the visualization requirements informed these decisions.

\subsubsection{Topic Modeling}
Starting with the synthetic dataset, the system generates an embedding for each data item using OpenAI's ``text-embedding-3-small'' model. We first apply k-nearest neighbors (k-NN)~\cite{cover1967knn} with cosine similarity ($k=7$) to establish a stable, high-level topical framework. This choice of a fixed-count, centroid-adjacent method ensures a manageable ``birds-eye view'' of the dataset, preventing the fragmentation often seen in density-based methods at high dimensions. Once these broad thematic boundaries are set, we generate primary topic labels using GPT-4o following a similar approach in BERTopic~\cite{grootendorst2022bertopic}.

To resolve finer structures within these broad categories, we transition to a local clustering phase. We apply KernelPCA~\cite{scholkopf1997kernel} to project the embeddings of each primary cluster into 2D coordinates, reducing dimensionality while preserving non-linear relationships. We then perform HDBSCAN~\cite{mcinnes2017hdbscan} on these projections to identify subtopics. This two-step hybrid approach is motivated by the differing requirements of global vs. local discovery: while k-NN provides the structural stability needed for a consistent user interface, HDBSCAN’s density-based logic is better suited for discovering natural, irregularly shaped sub-clusters without forcing outliers into subtopics. This hierarchy effectively declutters the visualization (\autoref{sec: treemap}) by separating core thematic nodes from noise. Finally, subtopic labels are generated following the same procedure as the primary topics.



\subsubsection{Context Retrieval and Management}
At runtime, the context retrieval and management in VizCopilot follow that of M365 Copilot, but in a simplified form. The chatbot maintains a context block immediately following the system prompt. When the user sends the first prompt in a conversation, the chatbot retrieves context using a hybrid search that combines embedding similarity and keyword matching. The retrieved data items are converted to strings according to their data types and filled into the context block. Unless users explicitly modify the context via the data context panel, the context block remains unchanged throughout the conversation. Although this implementation is not technically on par with commercial products, it is sufficient to examine the viability of involving users in context engineering and to inform future enhancements, as we discuss in~\autoref{sec: user_study}.

\subsubsection{AI-generated Subtopic Summaries}
When the user enters a prompt, in addition to providing a direct response, VizCopilot generates summaries for each relevant subtopic, along with explanations of relevancy. These summaries are displayed in the visualization when users click on a subtopic tag. As with topic label generation, we check the length of the relevant items and apply a sampling strategy to prevent exceeding the context window of GPT-4o.

\subsection{Synthetic Dataset Generation}\label{sec: synthetic_dataset}
Due to privacy and security policies that prohibit access to other individuals' accounts, we did not use real enterprise data. Instead, we use a synthetic dataset that reflects the messy enterprise database. The dataset is about a fictitious AI companion company with about 1000 employees and around 10,000 items of enterprise data, including emails, files (pptx, docx, etc.), calendar events, and chat messages. Next, we introduce how the dataset is generated and discuss its limitations.

\subsubsection{Overall procedure}
The generation process begins with the creation of 1,000 distinct employees with the company background, each with diverse names, titles, and descriptions of personal backgrounds and job descriptions.
Emails, files, and calendar events are generated based on these employee profiles.
For example, an email is generated by first choosing two employees as sender and receiver, and then generating the content based on their employee profiles. 

As LLM-based chatbots rely on text fields to search and reason about data, we generate a ``content'' field of each file, which is a description in text form rather than in its original format (e.g., pptx). While such content may not already exist in enterprise databases, there exist mature technical solutions that can produce such content.
Since LLM-based chatbots process text inherently, the dataset remains representative of the structure of real-world enterprise data.

\subsubsection{Realistic flaws in the dataset}
The synthetic dataset contains several ``flaws'' that mirror characteristics of real-world enterprise data~\cite{mukherjee2004enterprise}. It allows duplicate items, such as repeated employee names, and may include inconsistencies or conflicts between data items, such as inconsistent terminologies or overloaded terms. Employees can appear in files unrelated to their job descriptions. The metadata may not make sense, e.g., a file is dated before the creator joined the company. While not exhaustive, the dataset remains broadly representative of the structure and informational properties of real-world enterprise data.

\subsubsection{Limitations}\label{sec: dataset_limitation}
While suited for our exploratory study, the generalizability of the dataset is limited.
The distribution of data types may not reflect that of a real enterprise database, and the overall scale of data produced by a company with 1,000 employees is likely much larger. The dataset lacks explicit connections between items beyond the involved employees. Chat or email threads do not include multi-turn conversations. For these limitations, the synthetic dataset is only considered appropriate for developing a research prototype. 

%% file: sections/05_user_study.tex
\section{User Study}\label{sec: user_study}
The goal of the user study was to explore the potential of providing users control over context engineering through data visualization. We wanted to understand if it is a viable approach, its potential advantages, disadvantages, and user perceptions. To this end, we conducted a qualitative study that enabled us to observe users interacting with the research prototype and engage them in a short post-session interview comparing VizCopilot with a chat-only interface.


\paragraph{Study Design}
Each participant interacted with both a chat-only Copilot, an interface modeled after M365 Copilot, and with VizCopilot.
The interface for the chat-only Copilot approximated the conversational functionalities of M365 Copilot but did not include agentic features, such as setting up meetings, as we focus only on information sensemaking and synthesis.
VizCopilot used the same conversational interface but augmented it with a visualization panel that afforded additional user interactions. This design ensured that the underlying chatbot capabilities remained identical across both conditions, so we could observe the effect of visualization.
Participants completed similar information synthesis tasks in both conditions using the synthetic dataset about an AI companion company (\autoref{tab: task_sets}). 

\paragraph{Recruitment} 
We recruited 14 participants with prior experience using M365 Copilot via email invitation, including product designers, software engineers, and researchers, at both junior and senior levels.
Each session lasted approximately one hour, and participants received \$40 as compensation. 

\paragraph{Procedure}
The study was conducted in person on a 16-inch Windows laptop. Participants were first introduced to the research background and procedure and provided informed consent, which took about 10 minutes. They then completed a randomly assigned task set using the chat-only Copilot first (15-20 minutes), then VizCopilot (15-20 minutes). The order was not counterbalanced since participants had prior experience with M365 Copilot already, and usability issues and concerns with M365 Copilot are known. Our primary focus was to identify the benefits and drawbacks of visualization qualitatively. Chat-only Copilot was meant to give participants a baseline to compare with.
Before interacting with VizCopilot, a brief (5-minute) tutorial was given to introduce its visual and interaction capabilities.

\paragraph{Data Analysis}
Participants were asked to think aloud during task completion. After completing both conditions, they filled out a questionnaire adapted from the Overreliance Risk Identification and Mitigation Framework~\cite{passi2022overreliance}, covering transparency, trust, confidence, sense of control, and verification. The questionnaire primarily served to contextualize and prompt reflection during subsequent semi-structured interviews and was not used for quantitative analysis. 
 The first author conducted qualitative thematic analysis~\cite{braun2019reflecting} 
 on the screen-capture recordings and the interview transcripts.
 The themes were refined through discussions with co-authors to enhance interpretive rigor.
The recruitment, procedure, data collection, and compensation were approved by the Institutional Review Board (IRB).


\begin{table}[t]
\centering
\caption{Task sets in the user study}
\small
\begin{tabular}{@{}l>{\raggedright\arraybackslash}p{0.36\columnwidth}|l>{\raggedright\arraybackslash}p{0.36\columnwidth}@{}}
\toprule
\multicolumn{2}{c|}{\textbf{Task set 1}} & \multicolumn{2}{c}{\textbf{Task set 2}} \\
\midrule
T1 & Summarize everything related to the product design
   & T1 & Summarize everything related to user feedback \\
T2 & Who is Liam Johnson?
   & T2 & Who is Aisha Patel? \\
T3 & What has been done in marketing?
   & T3 & What has been done in software development? \\
\bottomrule
\end{tabular}
\label{tab: task_sets}
\end{table}

%% file: sections/06_results.tex
\section{Results}
 In this section, we present findings from the thematic analysis.
Overall, VizCopilot was well-received by the participants for its thoughtful visual design: although the interface displayed dense information, participants did not feel lost during the tasks. The interactions were considered intuitive, and despite requiring more coordination between views than a typical UI, participants were able to use them fluidly. Next, we present each theme in more detail.

\subsection{Interactions with context in chat-only Copilot}
\theme{1.1}{Participants felt disconnections between prompts and responses}
When responses felt too generic or did not address their intent, participants expressed a need to examine the underlying context beyond citations, but the chat-only Copilot offered limited support. 
The lack of access to such information caused frustration and reduced trust, as one participant noted, \textit{``I feel utterly disconnected from the information I'm receiving. (P14)''}.
This finding supports our decision to display in VizCopilot the whole context the chatbot received, to answer the question \textit{``where the responses come from''}, i.e., the overarching background of the response captured by the retrieved context.

\theme{1.2}{Folk methods were invented to probe context in chat-only Copilot}
In the absence of UI to access context, participants developed informal strategies to probe the information chat-only Copilot relied on. These included asking the same question multiple times to check for consistency, or rephrasing queries with different keywords to observe variations in responses. Some reversed their prompting style, e.g., changing ``Tell me about marketing.'' to ``Do you know anything about marketing?'' Participants explained that these strategies aimed to bypass Copilot's abstractive summarization and access the ``raw data,'' i.e., the underlying context that shaped the response. Participants spent considerable time probing the context to obtain reliable answers for each task, but these methods were generally ineffective.

\theme{1.3}{Participants followed a structured flow with VizCopilot} 
In contrast, participants demonstrated a more structured interaction flow with VizCopilot. After submitting a prompt, they typically skimmed the response, consulted the data context panel or the highlighted visualization, and cross-referenced the context with the response. They could explore visualizations in depth by expanding subtopic summaries and verifying file content as needed. Most participants leveraged the drag-and-drop feature to refine context, reporting noticeable improvements. With explicit UI support for accessing and modifying context, participants avoided the folk methods seen in interactions with chat-only Copilot.

\begin{figure}[t]
    \centering
    \includegraphics[width=\columnwidth]{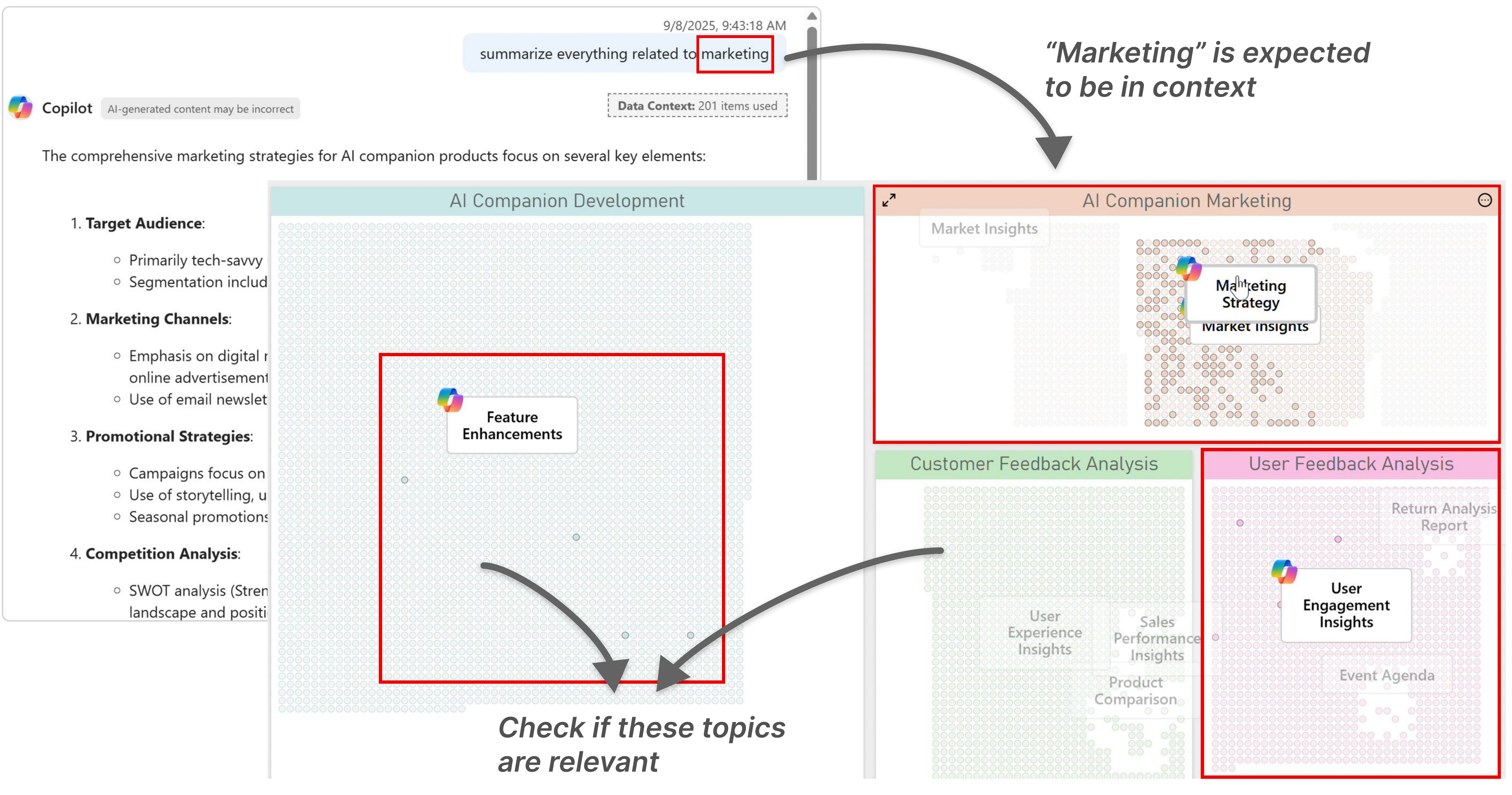}
    \caption{The highlight feature allows users to quickly check the alignment of retrieved context. When the user enters \textit{``Summarize everything related to marketing''}, the topic ``Marketing'' is expected to be highlighted, while the other two topics call for manual checks. }
    \Description{}
    \label{fig: transparency}
\end{figure}

\subsection{Improvements by Design in VizCopilot}
\theme{2.1}{Visibility and sensemaking support}
Participants appreciated the easy access to context in VizCopilot. As one participant noted, this helped them shape expectations about what VizCopilot should and should not retrieve: \textit{``I feel like I have a much better high-level view of what I'm actually asking and what my data actually looks like. I feel like I'm not shooting in the dark as much anymore.'' (P9)}  
After inputting a prompt, participants reported that glancing through the organized topics in the visualization allowed them to quickly sanity-check whether the retrieved context aligned with their intentions and to navigate toward areas of misalignment. The progressive levels of detail -- from topic labels, to AI-generated summaries, to file content -- helped ease the cognitive load of context sensemaking by enabling deeper inspection on demand, as illustrated by P3: \textit{``An overview is the most that I'd be looking at, and if the copilot starts hallucinating, I would want to go inside.'' (P3)}

\theme{2.2}{Increased sense of control}
In particular, participants were satisfied with the noticeable changes in responses when they tried to steer VizCopilot into their desired direction. Participants reported that fewer prompts were needed to get to their desired answer: \textit{``(VizCopilot is) much more efficient because then I wouldn't have to keep prompting the AI and adding context or moving context. …it helps align intention with the AI's interpretation of my prompt.'' (P3)''} 

\theme{2.3}{Visualization provides transparency of intermediate generative steps.}
The generative mechanism of chat-only Copilot, i.e., retrieving and synthesizing context to generate a response, was often unclear to participants. Making intermediate steps of this mechanism visible in VizCopilot helped them reason about why and how responses might be wrong. As P11 explained: \textit{``It made it much clearer how it was getting the context in the first place. It straight up said it will start with this context and filter it down, ... clarified for me what it was doing behind the scenes.'' (P11)}  

\theme{2.4}{Identifying errors in VizCopilot: Missing Context and Misinterpretation of Context}
Participants were able to identify two major types of errors. The first was \textbf{missing or redundant context}. For instance (\autoref{fig: transparency}), if users issued the query \textit{``summarize everything related to marketing,''} but the topic ``AI Companion Marketing'' was not highlighted, they could immediately recognize missing context. Conversely, topics such as ``Feature Enhancements'' or ``User Engagement Insights'' were not directly related to marketing and appeared only sparsely highlighted, signaling redundancy. In both cases, participants could drill down to verify and adjust the context as needed.  
\begin{figure}[t]
    \centering
    \includegraphics[width=\columnwidth]{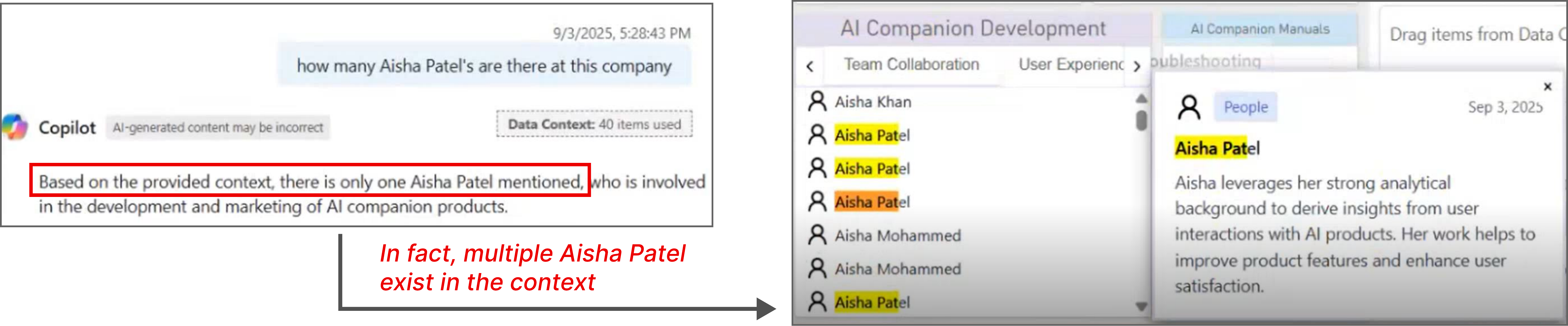}
    \caption{An example of Copilot misinterpreting the context. Copilot consistently mistakens different employees with the same name as the same person, despite the direct question. Most participants in the user study were able to identify such an error by inspecting the visualization panel in the file view.}
    \Description{}
    \label{fig: duplicate_names}
\end{figure}
 The second error was \textbf{misinterpretation of context}, as shown in~\autoref{fig: duplicate_names}. VizCopilot consistently conflated employees with duplicate names, treating them as the same person across both task sets. 
 This occurred because VizCopilot processes retrieved context in text form, potentially losing structure information. Although this error was harder to detect since the retrieved context initially appeared correct, 10 out of 14 participants were still able to identify it by simply glancing at the file view in the visualization. Identifying errors in AI outputs is essential for appropriate reliance. 

\theme{2.5}{Transparency increases trust and confidence}
Transparency also increased participants’ trust and confidence in the system. As P13 explained: \textit{``I can see where VizCopilot gets the data from to answer my questions, whether it could answer that question. I can tell if it's like hallucinating or something, so it gives me a little more trust in the AI.'' (P13)} 
Rather than interacting with a seemingly perfect black box, making the generative mechanism visible allowed participants to better judge the reliability of the information they received, which is essential for appropriate reliance.

\theme{2.6}{Prompting strategies are naturally adapted}
Most participants adapted their prompting strategy to help VizCopilot retrieve better context. For example, participants would reuse the topic labels as keywords and give more direct commands. Participants also liked how the visualization provided information for them to ask follow-up questions that yielded more specific answers: \textit{``"I spent a little bit of time looking (at the visualization), saw some product names, OK, I know there's different products, so now I need to write a prompt that tells me about what products they actually have.'' (P10)}
Our interpretation is that due to the visibility of context and transparency over the generative mechanism, participants found that their prompts are critical to the retrieval process and proactively adapted their prompting strategies. 

\subsection{Issues Remaining to be Solved}


\theme{3.1}{Better signals for verification are needed}
While context visualization helped support navigation to areas of interest in the context, participants were generally less likely to check responses when the prompts ``felt'' simple enough, which was highly subjective, inconsistent, and often a misperception. Better signaling supports, such as confidence or uncertainty scores, could be embedded into the interface and visualization design~\cite{stokes2024datauncertainty}. 

\theme{3.2}{User trust and close reading in verification needs better design}
Participants reported insufficient support once they reached the file view for verification. Although the AI-generated summaries for each subtopic were intended to reduce cognitive burden, some participants expressed distrust toward such summaries: \textit{``I just don't trust a summary. I would much rather it operate in such a way that it points me to exactly the file that would answer my question.'' (P14)} This highlights the continued need for designs that facilitate close reading, such as keyword highlighting, especially given that effectively facilitating verification is a critical strategy for mitigating overreliance~\cite{Microsoft_2025_OverrelianceAI}.

%% file: sections/07_discussion.tex
\section{Discussion}
Based on our findings, we reflect on the value of giving users control over context engineering, outline design implications, and present our vision for fostering appropriate reliance as well as transparent and sustainable human-AI interaction.
\subsection{Fostering appropriate reliance through visualization}
\subsubsection{Catching errors through visualization}
Prior works have investigated designs that leverage interventions such as uncertainty expressions~\cite{Kim2024uncertainty_expression} or cognitive forcing functions~\cite{buccinca2021cognitiveforcing, drosos2025itmakesthinkprovocations} to reduce overreliance on AI. Compared to these methods, visualization is a design that ``suggests'' rather than ``tells'' users about potentially incorrect answers. This creates a significantly different cognitive response in users, as they need to decide whether the issue (i.e., context misalignment) truly exists. While doing so, users engage in active sensemaking and critical reflection on the chatbot response, the context, and how they are connected. 
Theme 2.4 \textit{(Identifying errors in VizCopilot)} shows that users are capable of catching various AI errors, indicating that our approach has potential for fostering appropriate reliance. 


\subsubsection{Transparency through visualizaiton}
The capability of context visualization to foster appropriate reliance is in line with prior work that shows explanations and sources of an AI response can play a significant role in AI reliance~\cite{Kim2025explanations_sources_inconsistencies}. 
Here, explanations refer to supporting details or justifications for an AI answer. 
As opposed to AI-generated self-explanations, which are subject to unfaithfulness~\cite{si2024largelanguagemodelshelp}, 
context visualization makes the retrieved context visible and the retrieval process transparent without the concern of faithfulness.
By supporting users' understanding of how the AI chatbot derives the answer, our work follows prior work that calls for strategies that improve people's understanding of AI~\cite{Kim2025explanations_sources_inconsistencies, Annapureddy2025AI_literacy, nakao2022end_user_hil_AI}.
We show that with appropriate design, visualization has the potential to significantly reduce the cognitive load and technical barrier for understanding the retrieval process and reviewing the retrieved context in AI chatbots.

\subsubsection{Reducing anthropomorphic perceptions of AI through visualization}
Anthropomorphism of AI systems is one of the many concerns that influences appropriate reliance~\cite{kirk2023risk_taxonomy_llm, kramer2025trick_int_trust}. 
As context visualization provides transparency over the context retrieval process, it does not encourage anthropomorphization and appears to support realistic mental models. This is suggested by Theme 2.6 \textit{(Prompting strategies are naturally adapted)}, where participants learn to give more keywords and instructional prompts as they perceive the AI chatbot as simply a search-and-synthesis program. 
Moreover, visualization allows users to not solely rely on natural language interactions to steer the chatbot. This extra support of interaction and the increased sense of control (Theme 2.2) could also contribute to non-anthropomorphic perception.

\subsection{Design trade-offs: usability and appropriate reliance}
During the design of VizCopilot, we debated whether visualization should be displayed by default or shown on demand by collapsing the visualization panel. Several participants also raised the suggestion for on-demand during the user study. Our decision to present visualization by default rested on three considerations. 
First, context visualization supports sensemaking even before users enter a prompt; the topic structures are visually salient and provide an immediate high-level overview of available information.
Second, users often fail to recognize the need to verify responses~\cite{passi_vorvoreanu_aether2022overreliance}, a tendency that is exacerbated without a high-level overview of the context. Third, we envision that over time, users can develop a mental model and efficiently look for information in the visualization, but an on-demand design could hinder this learning process.

This design decision makes VizCopilot overwhelming at first encounter. Our research prototype incorporates intricate mechanisms that are rarely seen in enterprise products, such as 
a progressive visualization panel, and the coordination between the visualization and chat panel. Consequently, VizCopilot may suffer from low discoverability without guided orientation. 

We view this as a trade-off between immediate usability and the long-term goal of fostering appropriate reliance. 
We believe that immediate usability can be improved through well-designed onboarding support, and that further opportunities exist to enhance usability by integrating familiar search-oriented interactions into the visualization panel, such as search bars and metadata filters that enterprise users already recognize from existing tools. Although a visualization interface may impose a higher initial learning cost, it can help users develop more effective cognitive strategies that foster appropriate reliance on AI chatbots. Future research can further explore this trade-off.

\subsection{Sustainable human-AI collaboration}
VizCopilot speaks to the advantage of augmenting human capability rather than automating it~\cite{heer2019sharedrepresentation}. Once past the initial learning curve, users gradually develop a ``data sense'' of their work-related information, akin to the data hunches on uncertainty observed in prior studies~\cite{Lin2023data_hunches}. This capability enables them to navigate and locate task-relevant contexts while identifying misaligned ones, thereby empowering higher agency.

In contrast, the current AI system landscape, including conversational AI, has been widely criticized for its adverse impacts on human cognition and skills. Studies show that the language-based collaboration paradigm imposes significant metacognitive demands~\cite{tankelevitch2024metacognition} and can hinder critical thinking skills~\cite{lee2025critical_thinking_genai}. 
These findings raise concerns about the long-term consequences of AI.

These concerns were anticipated in the well-known debate on direct manipulation interfaces (represented by visualizations) versus software agents~\cite{shneiderman_maes1997} between Ben Shneiderman and Pattie Maes. VizCopilot integrates visualization and AI to balance user control and autonomous support, exemplifying the design that Shneiderman and Maes ultimately converged on. 
While AI technologies have advanced considerably, our work affirms that incorporating visualizations into AI systems can enhance human cognition and decision-making by offloading critical yet cognitively demanding tasks to visual representations. In doing so, it facilitates a sustainable paradigm of human–AI collaboration in which human agency is preserved and strengthened over time.

\subsection{Limitations}
This work has several limitations, primarily concerning the user study and the technical maturity of VizCopilot. First, the use of a synthetic dataset and research prototype limits us from fully capturing the real-world user experience, posing threats to ecological validity. This study demonstrates the potential value of a visualization-enabled approach to context engineering, but does not provide sufficient evidence to inform at-scale product design without further investigation. Second, participants had at most one hour to use VizCopilot, which is insufficient to assess value over time and potential long-term effects. Third, the unit visualization we used has scalability limits. We do not claim that using a treemap and unit visualization design is the best choice, but this study nonetheless demonstrates that leveraging corpus visualization designs can significantly increase the amount of visible data, which is critical for context alignment and ultimately fostering appropriate reliance in enterprise chatbots. We hope that our insights encourage future designers to further improve the visualization design to accommodate more data items or to generalize to more specific enterprise usage and databases if needed.

As a proof-of-concept system, the computational and visual scalability must be improved for larger datasets. The algorithm choices and hyperparameters also require tuning to generalize across domains. Moreover, VizCopilot's context retrieval mechanism is substantially simpler than those in commercial products.
VizCopilot is a research prototype, and while its limitations constrain the strength and generalizability of our conclusions, it helps establish the potential value of giving users direct control of context engineering and of enhancing enterprise AI chatbots with visualization. Having shown the value of this approach, this study provides a foundation for future research exploring and refining the use of context engineering and data visualization for synthesizing large amounts of enterprise data.

\section{Conclusion}
 Our study shows that giving users control over context engineering has benefits and that well-designed visualization can play a critical role in enabling them to align context for chatbots. Even as LLM-based systems advance and retrieval processes become more complex, the benefits of visualization we observed are likely to remain essential for sustainable human-AI collaboration. Our findings suggest that visualization should be considered a valuable design strategy in the development of future systems.